\documentclass[twocolumn]{autart}    
\usepackage{cite}
\usepackage{psfrag}
\usepackage{graphics}
\usepackage{epsfig, latexsym, amsmath, color, amsfonts, amssymb,graphicx}
\usepackage{algorithm}
\usepackage{pdfsync}
\usepackage{booktabs}
\usepackage{algpseudocode}
\usepackage{amsmath}
\usepackage{graphics}
\usepackage{epsfig}
\usepackage{caption}
\usepackage{subfigure}

\newtheorem{theorem}{Theorem}[section]
\newtheorem{lemma}[theorem]{Lemma}

\newtheorem{proposition}[theorem]{Proposition}
\newtheorem{remark}[theorem]{Remark}

\newtheorem{example}[theorem]{Example}

\newtheorem{problem}[theorem]{Problem}

\newcommand{\R}{{\rm  I\kern-2pt R}}
\renewcommand{\Re}{{\rm  I\kern-2pt R}}

\begin{document}

\begin{frontmatter}

\title{{Data-Driven Power Control for State Estimation:~A
Bayesian Inference Approach}\thanksref{footnoteinfo}}

\thanks[footnoteinfo]{The work of J.~Wu, Y.~Li and L.~Shi is
supported by a HK RGC GRF grant 618612. }

\author[Paestum]{Junfeng Wu}\ead{jfwu@ust.hk},
\author[Paestum]{Yuzhe Li}\ead{yliah@ust.hk},
\author[Rome]{Daniel E. Quevedo}\ead{dquevedo@ieee.org},
\author[Paestum]{Vincent Lau}\ead{eeknlau@ust.hk},
\author[Paestum]{Ling Shi}\ead{eesling@ust.hk}

\address[Paestum]{Department of Electronic and Computer Engineering, Hong Kong University of Science and Technology, Hong Kong.}
\address[Rome]{School of Electronic Engineering and Computer Science, University of Newcastle, Australia.}


\begin{abstract}
We consider sensor transmission power control for state
estimation, using a Bayesian inference approach.
A sensor node sends its local state estimate to a remote estimator over an
unreliable wireless communication channel with random data packet drops. As related to packet dropout rate, transmission power is chosen by the
sensor based on the relative importance of the local state estimate. The
proposed power controller is proved to preserve Gaussianity of
local estimate innovation, which enables us to obtain a closed-form
solution of the expected state estimation error covariance. Comparisons with
alternative non-data-driven controllers demonstrate performance
improvement using our approach.

\end{abstract}

\end{frontmatter}


\section{Introduction}\label{sec:intro}
{Wireless networked systems have a wide spectrum of applications in smart grid, environment monitoring,
intelligent transportation, etc.
State estimation is a key enabling technology where
the sensor(s) and the estimator communicate over a wireless network.
 Energy conservation is a crucial issue as most wireless sensors use on-board
 batteries which are difficult to replace and typically are expected to work for years without replacement.
Thus power control becomes crucial.
In this work, we consider sensor transmission power control for remote state
estimation over a packet-dropping network.
{Transmission power control in state estimation scenario has been considered
from different perspectives. Some works took transmission costs as constant.}
Shi \textit{et al.} \cite{shi2012optimal}
assumed sensors to have two energy
modes, allowing it to send
data to a remote estimator over an unreliable
channel either using a high or low transmission power
level. The optimal power controller is
to minimize the expected terminal estimation error at the remote estimator subject to an energy
constraint.
Similar works can also be found in~\cite{xu2004optimal,Imer05CDC}.
Meanwhile, some literature has taken channel conditions
into account.
Quevedo \textit{et al.}~\cite{queahl10} studied state estimation over fading channels.
They proposed a predictive control algorithm,
where power and cookbooks are determined in an online fashion based on
the undergoing estimation error covariance and the channel gain predictions.
{More related works can been
 seen in~\cite{leong2012power,nouleo14,quevedo2012kalman}.}

{An important issue which has not been taken seriously in most
works is that the transmission power assignment, as a tool to
control the accessibility of information to the receiver,
should be determined not only by the underlying channel condition and
the desired estimation performance, but also by the transmitted
information itself.
In~\cite{queahl10} and~~\cite{leong2012power},
the authors failed to associate transmission power with data to be sent.
The plant states are used to determine the transmission power in~\cite{GatsisACC13}.
In this case, lost packets signal the receiver of the state information. To avoid computation difficulty, the signaling information is discarded.}

In this paper, we focus on how to adapt the transmission power to the
measurements of plant state and how to exploit information contained in the lost packets. We propose a data-driven power controller, which utilizes
different transmission power levels to send the local estimate according to a quadratic function of a key parameter called ``incremental
innovation" which is evaluated by the sensor at each time slot.
By doing this, even when data dropouts occur, {the
remote estimator can utilize the additional signaling information
to refine the posterior probability density} of the estimation error by a Bayesian inference technique (see
\cite{box2011bayesian}), therefore deriving the
MMSE estimate. It compensates the
deteriorated estimation performance caused by packet losses.
{To facilitate analysis, we assume that a baseline power controller has already been
established based on different factors with regard to different settings, such as
the requirement of estimation performance as in~\cite{shi2012optimal}
or the channel conditions as in~\cite{queahl10,quevedo2012kalman,leong2012power}.
We are devoted to developing a power controller that
embellishes this baseline controller by adapting the transmission power
to the measurements such that the averaged power
with respect to all
possible values taken by the measurements does not exceed that of the baseline power controller.
The proposed power controller, driven by online measurements,
can run on top of non-data-driven power controllers, which results in
hierarchical power control mechanisms.
Then extension to a time-varying power baseline is established in
Section~\ref{subsec:vary-power-in-time}.} Note that a related controller
was first proposed in~\cite{Liyuzhe13CDC}, but as a special case of the
controller in this work. The main contributions of the present work
are summarized as follows.
\vspace{-3mm}
\begin{enumerate}
\item We propose a data-driven power control
strategy for state estimation with packet losses, which adapts the transmission power to the
measured plant states.

\item We prove that the proposed power controller preserves Gaussianity of the local innovation. It simplifies derivation of the MMSE estimate and leads to a closed-form expression of the expected state estimation error covariance.

\item {We present a tuning method for parameter design.
Despite of its sub-optimality, the controller is shown to perform
{not worse} than an
alternative non-data-driven one.}
\vspace{-3mm}
\end{enumerate}
The remainder of this paper is organized
as follows. In
Sections \ref{sec:problem-setup} and~\ref{sec:transmission-power-control}, we give mathematical
models of the considered system and introduce the data-driven transmission power controller. In Section~\ref{section:analysis}, we
present the MMSE estimate at the remote estimator and
{a sub-optimal power controller that minimizes an upper bound of the remote
estimation error}. In Section~\ref{sec:simulation}, comparisons with
alternative non-data-driven controllers demonstrate performance
improvement using our approach. Section~\ref{sec:conculsion} presents concluding remarks.

\textit{Notation}: 
$\mathbb{N}$ (and $\mathbb{N}_+$) is the set of nonnegative
(and positive) integers. $\mathbb{S}_{+}^{n}$
is the cone of $n$ by $n$ positive semi-definite
matrices. For a matrix $X$, $\lambda_i(X)$ is the $i$th
smallest nonzero eigenvalue. We abuse
notations $\mathrm{det}(X)$ and $X^{-1}$, which are used, in case of a singular matrix $X$, to denote the pseudo-determinant and
the Moore-Penrose pseudoinverse.
$\delta_{ij}$ is the Dirac delta function, i.e., $\delta_{ij}$ equals to $1$ when $i=j$ and $0$ otherwise. The notation $\mathrm{pdf}(\mathbf{x},x)$
represents the probability density function (pdf) of a random variable $\mathbf{x}$ taking value at $x$.

\section{State Estimation using a Smart Sensor} \label{sec:problem-setup}
Consider a linear time-invariant (LTI) system:
\begin{eqnarray}
  x_{k+1} & = & Ax_k + w_k, \label{eqn:process-dynamics} \\
  y_k & = & Cx_k + v_k,  \label{eqn:measurement-equation}
\end{eqnarray}
where  $k\in \mathbb{N}$, $x_k \in \mathbb{R}^{n}$ is the system state vector at time $k$, $y_k\in \mathbb{R}^{m}$ is the measurement obtained by the sensor, the state noise $w_{k} \in\mathbb{R}^{n} $ and observation noise $v_k \in \mathbb{R}^{m}$ are zero-mean i.i.d. Gaussian noises with $\mathbb{E}[w_{k}w_{j}^\prime] =\delta_{kj}Q$ ($Q\succeq 0$), $\mathbb{E}[v_{k}(v_{j})^\prime] = \delta_{kj}R$ ($R \succ 0$), $\mathbb{E}[w_{k}(v_{j})^\prime] = 0 \; \forall j,k\in\mathbb{N}$. The initial state $x_0$ is a zero-mean Gaussian random vector with covariance $\Pi_0\succeq 0$ and is uncorrelated with $w_k$ and $v_k$. $(A, C)$ is assumed to be detectable and $(A, Q^{1/2})$ is assumed to be stabilizable. Furthermore, we assume $A$ is Hurtwitz.\footnote{{Since we focus on remote state estimation in this paper, for any practically working systems (to be monitored alone), $A$ has to be Hurwitz. Otherwise, the system state will go unbounded and there is no real sensing device which can track an unbounded state trajectory. Adding a control input to regulate the system state for an unstable $A$ and studying its associated stability issue will be beyond the scope of this paper and will be left as our future work.}}

\begin{figure}[htbp]
\centering
  \includegraphics[width=7cm]{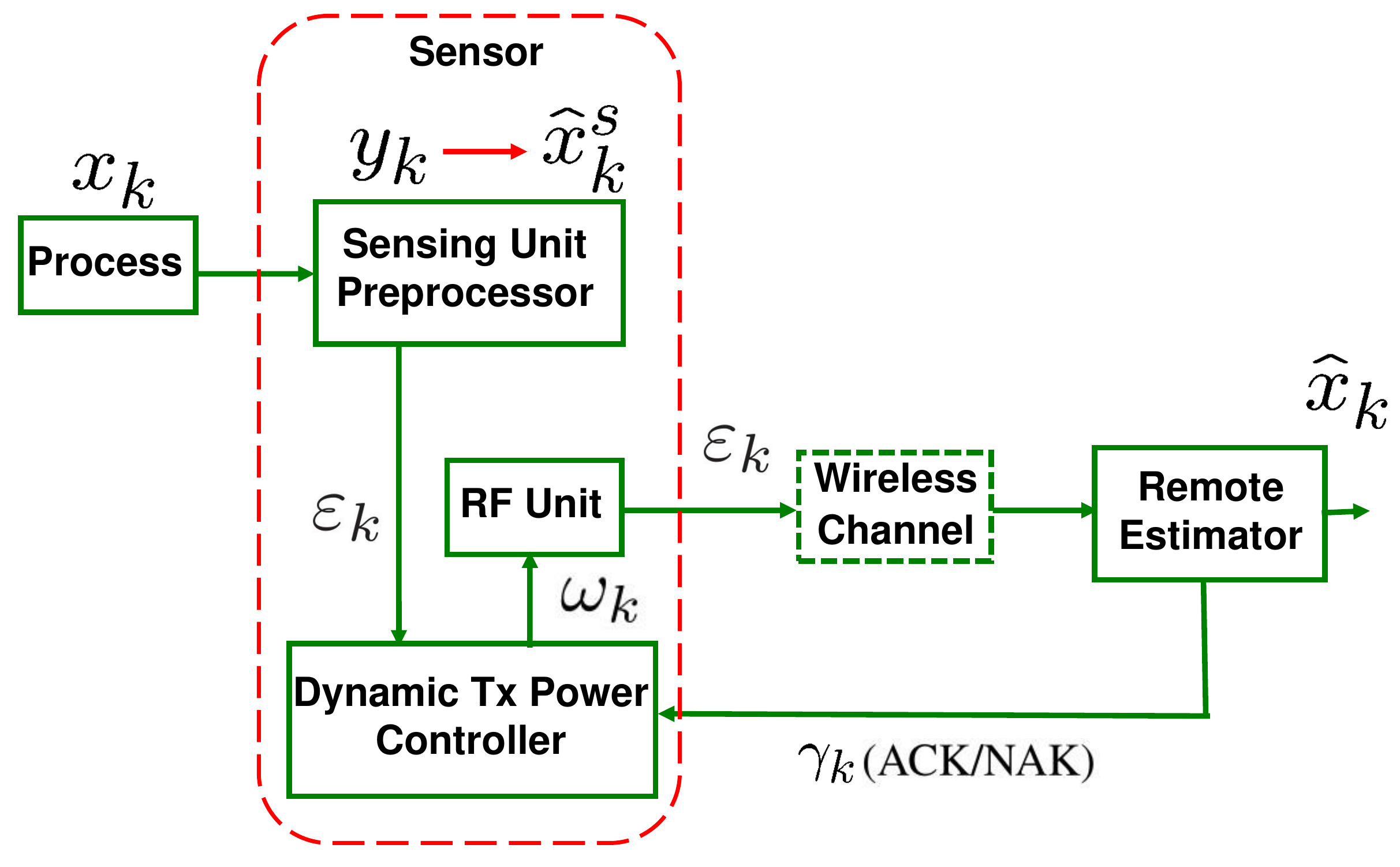}
  \caption{The system architecture.} \label{fig:system}
\end{figure}

\subsection{Sensor Local Estimate} \label{sec:local-state-estimate}
Hovareshti \textit{et al.}~\cite{hovareshti2007sensor} illustrated that utilization of the computation capabilities of wireless sensors may improve the system performance significantly.
Equipped with such
``smart sensors", 
the sensor locally runs a Kalman filter to produce the MMSE
estimate $\hat{x}_k^s$ of the state $x_k$ based on all the measurements
collected up to time $k$, i.e., $y_{1:k}\triangleq\{y_1,...,y_k\}$, and then
transmits its local estimate to the remote estimator.
Denote the sensor's local MMSE state estimate, the corresponding estimation error and error covariance as $\hat{x}_k^s$, $e^s_k$ and $P_k^s$, respectively, i.e., $\hat{x}_k^s\triangleq\mathbb{E}[x_k|y_{1:k}]$, $e_k^s\triangleq x_k-\hat{x}_k^s$ and ${P}_k^s\triangleq\mathbb{E}[(x_k-\hat{x}_k^s)(x_k-\hat{x}_k^s)'|y_{1:k}]$.
Standard Kalman filtering analysis suggests that
 these quantities can be calculated recursively (cf.,~\cite{andmoo79}),
where the recursion starts from $\hat{x}_{0}^s = 0$ and $P_0^s = \Pi_0\succeq0$.
Since $P_k^s$
converges to a steady-state value exponentially fast (cf.,~\cite{andmoo79}), we assume that the sensor's local Kalman filter has entered the steady state, that is,
$  P_k^s = \overline{P}\succeq 0~\forall k \in \mathbb{N}$,
{This assumption
simplifies our subsequent analysis and results, such as Theorem~\ref{theorem:form_P_k_on} and
Proposition~\ref{proposition: rank}.}
%

\subsection{Wireless Communication Model}

The data are sent to the remote estimator over
an Additive White Gaussian Noise (AWGN) channel using the Quadrature Amplitude Modulation
(QAM) whereby $\hat x_k^s$ is
quantized into $K$ bits and mapped to one of $2^K$ available QAM
symbols.\footnote{QAM is a common modulation scheme  widely used in IEEE 802.11g/n as well
as 3G and LTE systems, due to its high bandwidth efficiency.} 
For simplicity, the following assumptions are made:
\vspace{-3mm}
\begin{enumerate}
\item[A.1:]
The channel noise is independent of $w_k$ and $v_k$.

\item[A.2:]
$K$ is large enough so that quantization effect is negligible when
analyzing the performance of the remote estimator.
\item[A.3:]\label{asmpt:assumption-SER-2-gamma}
The remote estimator can
detect symbol errors\footnote{In practice, symbol errors can be detected via a cyclic redundancy check (CRC) code.}. Only the
data arriving error-free are regarded as being successfully received; otherwise they are regarded as dropout.
\end{enumerate}
\vspace{-3.5mm}
These assumptions are commonly used in communication and control
{theories~(cf.,\cite{sinopoli2004kalman,fu2009Automatica,queahl10,leong2012power,GatsisACC13}).}
{For example, Fu and Souza~\cite{fu2009Automatica} demonstrated that the estimation quality improvement (in terms of reduction of the remote estimation error) achieved by
increasing the number $K$ of the quantization bits is marginal when $K$ is sufficiently large (in their example $K$ only needs to be greater or equal to 4.} Based on A.3, 
the communication channel can be characterized by a random process $\{\gamma_k\}_{k\in\mathbb{N}_+}$, where\vspace{-1mm}
\begin{equation*} 
  \gamma_k =
  \begin{cases}
  1, &\text{if $\hat{x}^s_k$ arrives error-free at time $k$,}\\
  0, & \text{otherwise,}
  \end{cases}\vspace{-1mm}
\end{equation*}
initialized with $\gamma_0=1$.
{Denote $\gamma_{1:k}\triangleq \{\gamma_1,\ldots,\gamma_k\}.$}
Let $\omega_k\in[0,+\infty)$ be the transmission power for the QAM symbol
at time $k$.
We adopt the wireless communication channel model used in~\cite{Liyuzhe13CDC}, and have
${\mathrm{Pr}} \left(\gamma_k=0|\omega_k\right)=q^{\omega_k},$
where $q$ is given by $q\triangleq \exp( -\alpha /({N_0W}))\in(0,1),$
$N_0$ is the AWGN noise power spectral
density, $W$ is the channel bandwidth, and $\alpha\in (0,1]$ is a
constant that depends on the specific modulation being used.
To send local estimates to the remote estimator, the sensor chooses from a continuum of available power levels $\omega_k\geqslant 0$, see
Fig.~\ref{fig:system}. Note that different power levels lead to different
dropout rates, thereby affecting estimation performance.


\subsection{Remote State Estimation} \label{sec:remote-state-estim}
Define $\mathrm{I}_k$ as the information available to the remote estimator up to time $k$, i.e.,
\begin{equation}\label{eqn:I_k}
  \mathrm{I}_k=\{\gamma_1\hat{x}_1^s, \gamma_2\hat{x}_2^s,...,\gamma_k\hat{x}_k^s\}\cup
  \{\gamma_{1:k}\}.
\end{equation}
 Denote $\hat{x}_k$ and $P_k$ as the remote estimator's own MMSE state estimate and the corresponding estimation error covariance, i.e.,
  $\hat{x}_k\triangleq\mathbb{E}[x_k|\mathrm{I}_k]$ and
  ${P}_k\triangleq\mathbb{E}[(x_k-\hat{x}_k)(x_k-\hat{x}_k)'|\mathrm{I}_k]$,
{where expectations are taken with respect to a fixed
power controller.}
We
assume that 
the remote estimator feedbacks acknowledgements $\gamma_k$ before time $k+1$. Such setups are common especially when the remote
estimator (gateway) is an energy-abundant device.
This energy asymmetry 
allows the estimator to trade energy cost for estimation
accuracy.

\section{Data-driven Transmission Power Control}\label{sec:transmission-power-control}
Our strategy uses
the measurements  to assign
transmission power level efficiently.
{As focusing on how the power controller utilize the
sensor's real-time data,
to simplify discussion, we assume a constant power
baseline $\bar \omega$ in this section.
We define
$\theta\triangleq \{\theta_k\}_{k\in\mathbb{N}_+}$
as a transmission power controller over the entire time horizon, where
$\theta_k$ is a mapping from $y_{1:k}$ and $\gamma_{1:k}$ to $\omega_k$.
Before proceeding to study $\theta$, let us first briefly explain the idea of data-driven power control mechanism.
Define $\tau(k)\in\mathbb N_+$ as the holding time since
the most recent time when the remote estimator received the data from the sensor, i.e.,
\begin{equation} \label{definiton_tau}
  \tau(k)\triangleq k-\max_{1\leqslant t \leqslant k-1}\{t : \gamma_t=1\}.
\end{equation}
We interchange $\tau(k)$ with $\tau$ when the underlying time index is clear from the context.
Define $\varepsilon_k$ as the incremental innovation in the sensor local state estimate compared to time $k-\tau$, the previous reception instant, i.e.,
\begin{equation} \label{definition:z_k}
\varepsilon_k=\hat x_k^s-A^\tau\hat x_{k-\tau}^s.
\end{equation}
\begin{lemma} \label{lemma:z_k}
$\mathbb{E} [e_k^s\varepsilon_{k}'|\mathrm{I}_{k-1},\gamma_k=0]=0~\forall k\in \mathbb{N}_+.$
\begin{pf*}{Proof:}
{
The result follows from noting that
\begin{eqnarray*}
\mathbb{E} [e_k^s\varepsilon_{k}'|\mathrm{I}_{k-1},\gamma_k=0]
&=&\mathbb{E} \left[\mathbb{E} [e_k^s\varepsilon_{k}'|y_{1:k},\gamma_{1:k}]|\mathrm{I}_{k-1},
\gamma_k=0\right]\\
&=&\mathbb{E} \left[\mathbb{E} [e_k^s|y_{1:k}]\varepsilon_{k}'|\mathrm{I}_{k-1},
\gamma_k=0\right]=0,
\end{eqnarray*}
where the second equality holds because $e_k^s$ is independent of $\gamma_{1:k}$, and
the last equality holds since $\mathbb{E} [e_k^s|y_{1:k}]=0.$
}
\end{pf*}
\end{lemma}
Note that, if $\varepsilon_k=0$, then the sensor generates a local estimate,
$\hat x_k^s$ identical to the prediction $A^\tau \hat x_{k-\tau}^s$. We would say that, for the remote estimator, the
 ``value" of information contained in $\hat x_k^s$ is null.
As $\varepsilon_k$ becomes larger,
$\hat x_k^s$ has an increasing drift
from the prediction $A^\tau \hat x_{k-\tau}^s$ and the importance of the sensor sending $\hat x_k^s$ thereby
raises. Motivated by these observations, we define a stationary power controller, $\theta_{\rm ef}:\varepsilon_k\rightarrow \omega_k$,
as an increasing function
of $\varepsilon_k$. To fit the above observations, we introduce a quadratic function of $\varepsilon_k$ given by
$\mathcal{C}(\varepsilon_k,\mathcal{Q})\triangleq {\varepsilon_k}'\mathcal{Q}\varepsilon_k,$
where $\mathcal{Q}\in \mathbb{S}_+^n$ is a weight matrix.
According to~Lemma~\ref{lemma:z_k},
the covariance of $\varepsilon_k$ is a function of $\tau(k)$.
Therefore we specify $\tau(k)$ for the index
of $\mathcal{Q}$
and construct the following controller:
\begin{equation}\label{def:quadratic_schedule}
  \theta_{\text{ef}}: \{\omega_k=\frac{N_0W}{2\alpha}
  \mathcal{C}(\varepsilon_k,\mathcal{Q}_\tau)+
  \omega\}.
\end{equation}
In contrast to~\eqref{def:quadratic_schedule}, most non-data-driven transmission power controllers
(i.e.,~\cite{queahl10,leong2012power})
use a given power $\bar \omega$
regardless of what value $\varepsilon_k$ takes.
Note that in~\eqref{def:quadratic_schedule} a constant term $\omega$ is added
after $\mathcal{C}(\varepsilon_k, \mathcal{Q}_\tau)$.
If one sets $\mathcal{Q}_\tau=0$, then the transmission
with the baseline power controller $\omega=\bar \omega$ is a special case of the proposed
transmission power controller.
As for $\mathcal{Q}_\tau\not=0$,
the transmission power is a constant $\omega$ if $\mathcal{C}(\varepsilon_k \mathcal{Q}_\tau)= 0$; otherwise it is adapted according to $\mathcal{C}(\varepsilon_k,\mathcal{Q}_\tau)$.
{Compared with a related controller proposed earlier
in~\cite{Liyuzhe13CDC}, $\theta_{\text{ef}}$ in~\eqref{def:quadratic_schedule} is more general at least
from two aspects: 1) we introduce a weight matrix $\mathcal{Q}_\tau$ to highlight
the roles of different entries of $\varepsilon_k$; 2) it allows
the sensor to transmit using a standard power $\omega$ even
if $\mathcal{C}(\varepsilon_k, \mathcal{Q}_\tau)= 0$, which
includes a non-data-driven power transmission as a special case.}
As shown later in Lemma~\ref{lemma:still-gaussian}, given
$\mathrm{I}_{k-1}$,
$\varepsilon_k$ is zero-mean Gaussian with a
covariance $\Sigma_\tau$ depending on $\tau(k)$, i.e.,
$(\varepsilon_k|\mathrm{I}_{k-1})\sim
\mathcal{N}(0,\Sigma_{\tau}).$
For convenience of our subsequent analysis, we define a new parameter $\Psi_\tau$ satisfying
$\Psi_\tau\triangleq \left(\mathcal{Q}_k+\Sigma_\tau^{-1}\right)^{-1},$
where $\Sigma_\tau\succeq\Psi_\tau\succeq0$.
We now list the main
problems considered in the remainder of this work,
\begin{enumerate}
\item Under $\theta_{\rm ef}$ defined in~\eqref{def:quadratic_schedule}, what is the
    MMSE estimate and its associated
    estimation error covariance?
\item What value should $\mathcal{Q}_\tau$ (or $\Psi_\tau$) take in order
to minimize, $\mathbb{E}[P_k]$, the expected estimation error at the remote estimator?
\end{enumerate}
The solution to the first problem is presented in
Section~\ref{subsec:bayesian-reference-power-schedule}.
A sub-optimal solution to the second one is given
in Section~\ref{subsec:opt-bayesian-reference-schedule} in
view of the difficulty of the optimization problem.

{Before proceeding, we note that in previous works such as~\cite{GatsisACC13}
the difficulty of
using the information contained in lost packets, i.e., $\gamma_k = 0$, when computing
the MMSE estimate of the plant state has been acknowledged. One typically discards such information as was done in~\cite{GatsisACC13} or resorts to approximations, e.g., treating a truncated Gaussian distribution as a Gaussian distribution as was done in~\cite{wu2013event}. These approaches either lead to conservative results (due to the unutilized information) or inaccurate results (due to approximations).
Our method, on other hand, makes use of the information contained in the event $\gamma_k=0$ to improve the estimation performance. The associated MMSE estimate, relying on no approximation techniques, is derived in a closed-form.}

\section{Main Results} \label{section:analysis}

\subsection{Preliminaries}\label{subsec:preliminaries}

For any $\Sigma\succeq 0$ that is singular,
there exist
matrices $U,D\in\mathbb{R}^{n\times n}$
such that
$
\Sigma=UDU',
$
where $U$ is unitary, whose columns
are right eigenvectors of $\Sigma$,
and $D\triangleq\left[
\begin{array}{cc}
\Delta  & 0\\
0 & 0
\end{array}\right]$, where $\Delta$
is a diagonal matrix generated by the
corresponding nonzero eigenvalues
of $\Sigma$. Let $\Sigma^{1/2}\triangleq
U\sqrt{D}$. Then $\Sigma=\Sigma^{1/2}
\left(\Sigma^{1/2}\right)'.$

Generally speaking, an $n$-dimensioned random vector
$\mathbf{x}\sim \mathcal{N}(\mu, \Sigma)$,
does not have a pdf
with respect to the Lebesgue measure
on $\mathbb{R}^{n}$
if some entries in $\mathbf{x}$ degenerate
to almost surely constant random variables.
To work with such vectors, one can instead consider Lebesgue measure in the $\mathrm{rank}(\Sigma)$-dimension affine subspace: $\Omega\triangleq\{\mu+\Sigma^{{1}/{2}}\mathbf{z}:~
\mathbf{z}\in\mathbb{R}^n\}$, with respect to which
$\mathbf{x}$ has a pdf
$\mathrm{pdf}(\mathbf{x},x)
=
\frac{1}{\sqrt{\sigma}}\exp{
\left(-\frac{1}{2}(x-\mu)'
\Sigma^{-1}(x-\mu)
\right)},$
where $\sigma=(2\pi)^{\mathrm{rank}(\Sigma)}\mathrm{det}(\Sigma)$.
Without loss of generality, in the remainder of this paper,
for a random variable
$\mathbf{x}\sim \mathcal{N}(0,\Sigma)$ with a singular $\Sigma$, the pdf of $\mathbf{x}$ means
the probability density on $\Omega$.
Note that the Moore-Penrose
pseudoinverse of $\Sigma$ is unique and given
by
\begin{equation}\label{eqn:eigen-decompostion}
\Sigma^{-1}=U\left[
\begin{array}{lc}
\Delta^{-1} & 0\\
 0&0\end{array}\right]U',
\end{equation}
and that the
pseudo-determinant of $\Sigma$ equals to the
product of all nonzero eigenvalues of $\Sigma$.

Consider the power control law $\theta_{\rm ef}$ defined in~\eqref{def:quadratic_schedule}.
In order to guarantee that $\omega_k$ is always nonnegative for any value $\varepsilon_k$, the difference of
$\Psi_\tau^{-1}$ and $\Sigma_\tau^{-1}$ needs to
be at least positive semi-definite, i.e.,
two conditions must be simultaneously satisfied, which
are  $\Sigma_\tau\succeq\Psi
_\tau$ and $\Psi_\tau^{-1}\succeq\Sigma
_\tau^{-1}.$ The following lemma
provides a necessary condition that $\Psi_\tau$ needs to satisfy.
\begin{lemma}\label{lemma:sigma-psi-property}
Suppose $\Sigma$ and $\Psi$ satisfy $\Sigma\succeq\Psi
$ and $\Psi^{-1}\succeq\Sigma^{-1}$. Then
\begin{equation}\label{eqn:idential-rank1}
\mathrm{rank}(\Psi)=
\mathrm{rank}(\Sigma)
\end{equation}
and
\begin{equation}\label{eqn:same-image}
\mathrm{Im}(\Sigma^{{1}/{2}})=
\mathrm{Im}(\Psi^{{1}/{2}}),
\end{equation}
where $\mathrm{Im}(X)$ is the image of $X$.
\begin{pf*}{Proof:}
Since $\Sigma\succeq \Psi$, it is true that
$\mathrm{rank}(\Sigma)\geq
\mathrm{rank}(\Psi)$.
To verify \eqref{eqn:idential-rank1}, suppose that $\mathrm{rank}(\Sigma)>
\mathrm{rank}(\Psi)$.
Then from~(\ref{eqn:eigen-decompostion}), $\mathrm{rank}(\Sigma^{-1})>
\mathrm{rank}(\Psi^{-1})$,
which contradicts with $\Psi^{-1}\succeq\Sigma^{-1}$.
To prove~\eqref{eqn:same-image}, let us
denote $\mathrm{rank}(\Psi)\triangleq r$ and  assume there is a set of vectors $\mathbf{W}\triangleq\{\mathbf{w}_1,\ldots,
\mathbf{w}_r\}$
such that
$\mathrm{Im}(\Psi^{{1}/{2}})=
\mathrm{span}\left(
\{\mathbf{w}_1,\ldots,\mathbf{w}_r\}
\right).$
Suppose $\mathrm{Im}(\Sigma^{{1}/{2}})
\neq\mathrm{Im}(\Psi^{{1}/{2}})$. Then
there exists a vector in
$\mathbf{W}$ (without
loss of generality, let it be $\mathbf{w}_1$),
and a vector $\mathbf{w}_0
\in\mathrm{Ker}
\left((\Sigma^{1/2})'\right)$
where the operator $\mathrm{Ker}(X)$ is the kernel of a matrix $X$, such that ${\mathbf{w}_0}'
\mathbf{w}_1\neq 0$.
It leads to the fact that
$\mathbf{w}_0\not\in \mathrm{Ker}
\left((\Psi^{1/2})'\right)$.
We in turn have
$${\mathbf{w}_0}'\Sigma^{{1}/{2}}
\left(\Sigma^{{1}/{2}}\right)'
\mathbf{w}_0=0\hbox{~~while~~}
{\mathbf{w}_0}'\Psi^{{1}/{2}}
\left(\Psi^{{1}/{2}}\right)'
\mathbf{w}_0>0,
$$
which contradicts with $\Sigma\succeq\Psi$.
\end{pf*}
\end{lemma}
For convenience, denote
$n_\tau\triangleq \mathrm{rank}(\Sigma_\tau)=\mathrm{rank}
(\Psi_\tau)$, $\Omega_\tau\triangleq \mathrm{Im}
(\Sigma_\tau^{1/2})=\mathrm{Im}
(\Psi_\tau^{1/2})$ and
$\Phi_\tau\triangleq
\left(\Sigma_\tau^{1/2}\right)'
\Psi_\tau^{-1}\Sigma_\tau^{1/2}.$
One has next lemma, the proof provided in the Appendix.
\begin{lemma}\label{lemma:rank-Phi}
The rank of $\Phi_\tau$ equals that of $\Sigma_\tau$ (or $\Psi_\tau$), i.e.,
$
\mathrm{rank}(\Phi_\tau
)=n_\tau.
$
\end{lemma}
\begin{example}
Two matrices are provided below
as a simple example for $n=3$,
$$
\Sigma_\tau=\left[\begin{array}{ccc}
5 & 0 &0\\
0&5&0\\
0&0&0\end{array}\right], \hbox{~~and~~}
\Psi_\tau=\left[\begin{array}{ccc}
3 & -1 &0\\
-1&3&0\\
0&0&0\end{array}\right].
$$
We can verify that $n_\tau=2$, $\Sigma_\tau\succeq \Psi_\tau$,
$\Psi_\tau^{-1}\succeq \Sigma_\tau^{-1}$,~(\ref{eqn:idential-rank1}), and Lemma~\ref{lemma:sigma-psi-property} holds.
\end{example}


\subsection{MMSE State Estimate}\label{subsec:bayesian-reference-power-schedule}

In general, the posterior distribution of $\varepsilon_k$ fails to maintain Gaussianity without
analog-amplitude observations. The defect is
especially common for quantized Kalman filtering and Gaussian filters, where
it is tackled by Gaussian approximation \cite{andmoo79,kotecha2003gaussian,soi-kf-tsp06}.
By contrast,
the following lemma shows that,
using $\theta_{\rm ef}$ in \eqref{def:quadratic_schedule},
the distribution of $\varepsilon_k$ conditioned on
$\mathrm{I}_{k-1},\gamma_k=0$
is Gaussian. The proof, similar to that of Lemma 3.5 in~\cite{Liyuzhe13CDC}, is omitted.
\begin{lemma} \label{lemma:still-gaussian}
Under $\theta_{\rm ef}$ defined in~\eqref{def:quadratic_schedule}, given $\mathrm{I}_{k-1}$,
$\varepsilon_{k}$ follows a Gaussian distribution:
$ (\varepsilon_k|
  {\mathrm{I}_{k-1}}) \sim \mathcal
  N(0, {\Sigma_\tau}),$
where $\Sigma_\tau$ is given by the following recursion:
\begin{equation}\label{eqn:evolution_sigma_psi}
\Sigma_\tau=A\Psi_
{\tau-1}A'+\left(h(\overline{P})-\overline{P}\right),
\end{equation}
with $\Psi_0=0$. It is also true that, given
$\gamma_k=0$ and $\mathrm{I}_{k-1}$,
$
  (\varepsilon_k|{\mathrm{I}_{k-1}},\gamma_k=0)
  \sim \mathcal N(0, {\Psi_\tau}).
$
%
%
%
\end{lemma}
\begin{proposition} \label{lemma:communication_rate}
Under $\theta_{\rm ef}$ defined in~\eqref{def:quadratic_schedule}, given $\mathrm{I}_{k-1}$,
the packet drop rate at time $k$ is given by $\mathrm{Pr}(\gamma_k=0|\mathrm{I}_{k-1})=\frac{1}{\sqrt{\mathrm{det}(\Sigma_\tau)
 \mathrm{det}(\Psi_\tau^{-1})}}
 \exp\left(-\frac{\alpha}{N_0W}\omega\right).$
\end{proposition}
We denote the packet arrival rate
as $p_\tau\triangleq1-\mathrm{Pr}(\gamma_k=0|\mathrm{I}_{k-1}),$ where
the subscript $\tau$ is to
emphasize that it depends on $\Sigma_\tau$ and $\Psi_\tau$.
To ensure that the averaged transmission power with respect to different values
taken by the measurement in $\theta$ does not exceed
$\bar \omega$, i.e.,
$\mathbb{E}[\omega_k|\mathrm{I}_{k-1}]\leq \bar\omega$, we require the following result.
\begin{lemma} \label{lemma:energy beta}
Under $\theta_{\rm ef}$~\eqref{def:quadratic_schedule}, given $\mathrm{I}_{k-1}$, the relation between $\mathbb E[\omega_k|\mathrm{I}_{k-1}]$ and $\Psi_\tau$, and $\omega$ is given by
\begin{equation}\label{eqn:expected-power}
\mathbb E[\omega_k|\mathrm{I}_{k-1}]=
\frac{N_0W}{2\alpha}\left(\mathrm{Tr}(
\Sigma_\tau\Psi_\tau^{-1})-{n_\tau}\right)+\omega.
\end{equation}
\begin{pf*}{Proof:}
From Lemma \ref{lemma:still-gaussian}, we know that
$ (\varepsilon_k|
  {\mathrm{I}_{k-1}}) \sim \mathcal
  N(0, {\Sigma_\tau}).$
Under $\theta_{\rm ef}$, we have:
\begin{eqnarray*}
  \mathbb E[\omega_k|\mathrm{I}_{k-1}]&=&\mathbb E\left[\mathbb E[\omega_k|\varepsilon_k]|\mathrm{I}_{k-1}\right]\\
  &=&\frac{N_0W}{2\alpha}\mathbb E\left[\varepsilon_k'
  \left(\Psi_\tau^{-1}-\Sigma_\tau^{-1}
  \right)
  \varepsilon_k\right|\mathrm{I}_{k-1}]+\omega\\
      &=&\frac{N_0W}{2\alpha}
   \mathrm{Tr}\left(\mathbb E\left[\varepsilon_k
  \varepsilon_k'|\mathrm{I}_{k-1}\right](\Psi_\tau^{-1}-\Sigma_\tau^{-1})
   \right)+\omega\\
  &=&\frac{N_0W}{2\alpha}\left(\mathrm{Tr}(
\Sigma_\tau\Psi_\tau^{-1})-{n_\tau}\right)+\omega.
\end{eqnarray*}
\end{pf*}
\end{lemma}
With $\theta_{\text{ef}}$ defined in \eqref{def:quadratic_schedule}, the remote estimator computes $x_k$ and $P_k$ according to the following two theorems.
\begin{theorem}\label{theorem:mmse-proof}
Under $\theta_{\mathrm{ef}}$~\eqref{def:quadratic_schedule},
the remote estimator computes $\hat{x}_k$ as
\begin{equation}\label{eqn:mmse-estimate}
\hat{x}_k = \left\{\begin{array}{ll}\hat{x}_k^s,
 & \mathrm{if}~\gamma_k = 1, \\
  A^{\tau}\hat x_{k-\tau}^s, & \mathrm{if}~\gamma_k = 0,\end{array}\right.
\end{equation}
where $\hat x_k^s$ is updated as $\hat x_k^s=A^{\tau}\hat x_{k-\tau}^s+\varepsilon_k$ when $\gamma_k=1$.
\begin{pf*}{Proof:}
    When $\gamma_k = 1$, the result is straightforward
    since $\hat{x}_k^s$ is the MMSE estimate of $x_k$ given
     $y_{1:k}$. Now consider $\gamma_k = 0$. 
   The tower rule gives
    \begin{eqnarray*}
    \mathbb{E}\left[x_k|\mathrm{I}_{k-1},
    \gamma_k=0\right]
        &=&\mathbb{E}\left[\mathbb{E}
    \left[x_k|y_{1:k},\gamma_{1:k}\right]
    |\mathrm{I}_{k-1},
    \gamma_k=0\right]\\
    &=&\mathbb{E}\left[A^\tau\hat x_{k-
    \tau}^s\hspace{-1mm}+\varepsilon_k|
    \mathrm{I}_{k-1},\gamma_k=0\right]\\
    &=&A^\tau\hat x_{k-\tau}^s+
    \mathbb{E}\left[\varepsilon_k|
    \mathrm{I}_{k-1},\gamma_k=0\right].
    \end{eqnarray*}
    Lemma~\ref{lemma:still-gaussian} leads to
    $\mathbb{E}\left[\varepsilon_k|
    \mathrm{I}_{k-1},\gamma_k=0\right]=0$.
    \end{pf*}
\end{theorem}
\begin{theorem} \label{theorem:form_P_k_on}
Under $\theta_{\mathrm{ef}}$~\eqref{def:quadratic_schedule},
$P_k$ at
the remote estimator is updated as
\begin{equation} \label{eqn:on-line-update-p}
    P_k = \left\{\begin{array}{ll}\overline{P}, &  \mathrm{if}~\gamma_k=1,\\
                            \overline P+\Psi_\tau ,  & \mathrm{if}~\gamma_k=0.
\end{array}\right.
\end{equation}
\begin{pf*}{Proof:}
    When $\gamma_k=1$ the result is straightforward.
    We only prove the case when $\gamma_k=0$.
    \begin{eqnarray*}
    &  &\mathbb{E}\left[(x_{k}-\hat{x}_{k})
    (x_{k}-\hat{x}_{k})'|\mathrm{I}_{k-1}, \gamma_{k}=0\right] \\
    &=&\mathbb{E}\left[(x_{k}-
    A^{\tau}
    \hat{x}_{k-\tau}^s)(x_{k}-
    A^{\tau}
    \hat{x}_{k-\tau}^s)'|
    \mathrm{I}_{k-1}, \gamma_{k}=0\right] \\
    & = & \mathbb{E}\left[\mathbb{E}\left[
    (e_k^s +\varepsilon_k)(\cdot)'
    |y_{1:k},\gamma_{1:k}\right]
    |\mathrm{I}_{k-1},\gamma_{k}=0\right]\\
    & = & \mathbb{E}[(e_{k}^s)(e_{k}^s)'|y_{1:k}]+
    \mathbb{E}\left[(\varepsilon_k)(\varepsilon_k)'|
    \mathrm{I}_{k-1},\gamma_k=0\right]\\
    & = & \overline{P} +\Psi_\tau,
    \end{eqnarray*}
    where the third equality is due to
    Lemma~\ref{lemma:z_k} and the last one
    is from Lemma~\ref{lemma:still-gaussian}.
\end{pf*}
\end{theorem}
\begin{remark}
{Under a baseline power controller with a constant power control $\bar \omega$, the remote estimator's estimate still obeys the recursion~\eqref{eqn:mmse-estimate}; however, the estimation error covariance is updated differently: $P_k=\overline{P}$ when $\gamma_k=1$, and $P_k=h(P_{k-1}) = \Sigma_{\tau}$ when $\gamma_k=0$). Note that although the obtained estimates under the two power controllers are the same, their different estimation error covariance matrices suggest different confident levels with which the remote estimator trusts the obtained estimate: with the data-driven power controller, it is more convinced that the obtained estimate is close to the real state while less convinced with a non-data-driven power controller.}
\end{remark}
%
%
\subsection{{Selection of Design Parameters}}
\label{subsec:opt-bayesian-reference-schedule}
The performances of $\theta_{\rm ef}$ for different
$\Psi_\tau$'s are difficult to compare in general.
However, for $\Sigma_\tau$ and
$\Psi_\tau$, there must exist a
real number $\epsilon_\tau\in (0,1]$
such that $\Psi_\tau\preceq \epsilon_\tau\Sigma_\tau$ and
$\Psi_\tau\not\preceq \epsilon\Sigma_\tau,\forall~\epsilon<\epsilon_\tau$.
Observe that
$$
\Phi_\tau=\left(\Sigma_\tau^{1/2}\right)'\Psi_
\tau^{-1}\Sigma_\tau^{1/2}
\succeq \frac{1}{\epsilon_\tau}\left[
\begin{array}{cc}
I_{n_\tau} & 0\\
0& 0\end{array}\right],
$$
which yields
$\epsilon_\tau=\frac{1}
{\lambda_1(\Phi_\tau)}.$
In light of
(\ref{eqn:evolution_sigma_psi}),
we further have
$
\Psi_\tau\preceq
\epsilon_\tau\Sigma_\tau
=\epsilon_\tau\left(
A\Psi_{\tau-1}A'+\Sigma_1
\right).$
According to Proposition~\ref{lemma:communication_rate},
it can be seen given $\tau(k)=\tau$ that $\mathbb{E}[P_k|\tau(k)=\tau]$ has an
upper bound:
$\mathbb{E}[P_k|\tau(k)=\tau]\preceq
\overline{P}+(1-p_\tau)\epsilon_\tau
\left(
A\Psi_{\tau-1}A'+\Sigma_1
\right).$
Instead of minimizing $\mathbb{E}[P_k]$,
we minimize its upper bound which is equivalent to
minimize $(1-p_\tau)\epsilon_\tau$. Iterating
over time, one eventually needs to minimize $(1-p_{\tau})\epsilon_{\tau}$ for
any $\tau(k)\in \mathbb{N}_+$ at any $k\in \mathbb{N}_+$.
To this end, we propose to assign parameters of $\theta_{\rm ef}$ in
\eqref{def:quadratic_schedule} as the solution to the following
optimization problem:
\begin{problem}\label{problem:optimization_online}
  \begin{eqnarray*}
  & & \min_{\Psi_
    \tau,\Sigma_\tau,\omega}~~\frac{1}
    {\left(\mathrm{det}(\Sigma_\tau)
    \mathrm{det}(\Psi_\tau^{-1})\right)^{1/2}\hspace{-1mm}
    \lambda_1(\Phi_\tau)}\exp{\hspace{-1mm}\left[-\frac{\alpha}{
    N_0W}\omega\right]},\\
    & & ~~\,\mathrm{s.t.} ~~~~~
    \frac{N_0W}{2\alpha}\left(
    \mathrm{Tr}(\Sigma_\tau\Psi_\tau^{-1})-n_\tau\right)+\omega\leq \bar \omega.
  \end{eqnarray*}
\end{problem}
The constraint is imposed by~(\ref{eqn:expected-power}). To solve Problem~\ref{problem:optimization_online}, we
first note that
$\mathrm{Tr}(\Sigma_\tau
\Psi_\tau^{-1})=\mathrm{Tr}
(\Phi_\tau).$
However, for any matrix $X,Y\in \mathbb{R}^{n\times n}$,
$\mathrm{det}(XY)=\mathrm{det}
(X)\mathrm{det}(Y)$ does not hold in general since $\mathrm{det}(X)$ means $X$'s
pseudo-determinant (in case $X$ is singular). Fortunately, this property still holds for $\Sigma_\tau$ and $\Psi_\tau^{-1}$.
The proof is given in the Appendix.
\begin{lemma}\label{lemma:det-transform}
Suppose $\Sigma_\tau$ and $\Psi_\tau$
satisfy $\Sigma_\tau\succeq \Psi_\tau\succeq 0$ and
$\Psi_\tau^{-1}\succeq \Sigma_\tau^{-1}$.
Then $\mathrm{det}(\Sigma_\tau)\mathrm{det}(\Psi_\tau^{-1})
=\mathrm{det}(\Phi_\tau).$
\end{lemma}
From linear algebra,
$
\mathrm{det}(\Phi_\tau)=
\prod_{i=1}^{n_\tau}\lambda_i(\Phi_\tau)$,
and $
\mathrm{Tr}(\Phi_\tau)
=\sum_{i=1}^{n_\tau}\lambda_i(\Phi_\tau).
$
We simply write
$\lambda_i(\Phi_\tau)$ as $\lambda_i(\tau)$,
and denote the nonzero
eigenvalues of $\Phi_\tau$ by
$\Lambda_\tau\triangleq
[\,\lambda_{1}(\tau),\ldots,\lambda_
{n_\tau}(\tau)\,]$. Then
Problem~\ref{problem:optimization_online}
can be recast as
\begin{problem}\label{problem:transform_optimization_online}
      \begin{eqnarray}
    \min_{\Lambda_\tau,\omega}
    &&\frac{1}{
    \lambda_1(\tau)\prod_{i=1}^{n_\tau}
    \lambda_i(\tau)^{1/2}}
    \exp{\left[-\frac{\alpha}{
    N_0W}\omega\right]},\label{eqn:objective-func-opt}\\
     \mathrm{s.t.} &&
    \frac{N_0W}{2\alpha}\left[
    \sum_{i=1}^{n_\tau}\lambda_i(\tau)-
    n_\tau\right]+\omega= \bar \omega,~ \omega\geq 0\notag\\
    &&1\leq\lambda_1(\tau)\leq \lambda_j(\tau),~
    \forall j=2,\ldots,n_\tau.\notag
  \end{eqnarray}
\end{problem}
\begin{lemma}\label{lemma:equal-eig}
Let $\Lambda_\tau^*$ be the optimal solution to Problem~\ref{problem:transform_optimization_online}.
Then $\Lambda_\tau^*$ satisfies
\begin{equation}\label{eqn:opt-eig-equal}
\lambda_1(\tau)^*=\lambda_2(\tau)^*
=\cdots=\lambda_{n_\tau}(\tau)^*.
\end{equation}
\begin{pf*}{Proof:}
Suppose that
$\Lambda$
is the optimal solution to
Problem~\ref{problem:transform_optimization_online}
but does not satisfy~\eqref{eqn:opt-eig-equal}.
We will show that there must exist another
vector, which is different from $\Lambda$ and has
a smaller cost function~\eqref{eqn:objective-func-opt}.
Let $\sum_{i=1}^{n_\tau}\lambda_i=c$ where
$c$ is a positive constant.
Due to the fact that
$\lambda_1$ in $\Lambda$ is the minimum eigenvalue of
$\Phi_\tau$ and the inequality of arithmetic
and geometric means, we have
$\lambda_1\leq \frac{c}{n_\tau}$ and $\prod_{i=1}^{n_\tau}\lambda_i
\leq \left(\frac{c}{n_\tau}\right)^{n_\tau},$
the equalities simultaneously
satisfied when $\lambda_i=\frac{c}{n_\tau},~\forall~
i=1,\ldots,n_\tau$. Thus, $\Lambda_0=[
{\frac{c}{n_\tau},\ldots,
\frac{c}{n_\tau}}]$ results in a smaller value
of~\eqref{eqn:objective-func-opt}, which
contradicts with the assumption and
completes the proof.
\end{pf*}
\end{lemma}
The following lemma is a result of Lemma~\ref{lemma:equal-eig}. Its proof is
presented in the Appendix.
\begin{lemma}\label{thm:opt-sol}
If $\bar\omega>\frac{N_0W}{\alpha}$, then the optimal
solution to Problem~\ref{problem:transform_optimization_online}
is $\omega=\bar \omega-\frac{N_0W}{\alpha}$ and
\begin{equation}\label{eqn:opt_sol1}
\Lambda_\tau^{*}=
[1+\frac{2}{n_\tau},\ldots, 1+\frac{2}{n_\tau}].
\end{equation}
Otherwise, if $\bar \omega\leq \frac{N_0W}{\alpha}$, the optimizer is $\omega=0$ and
\begin{equation}\label{eqn:opt_sol2}
\Lambda_\tau^{*}=
[1+\frac{2\alpha\bar \omega}{n_\tau N_0W},\ldots, 1+\frac{2\alpha\bar \omega}{n_\tau N_0W}].
\end{equation}
\end{lemma}
Denote by $\theta_{\rm ef}^*$ the transmission power associated with
the solution to Problem~\ref{problem:transform_optimization_online}. Then we have the following theorem. It can be readily
verified from Lemma~\ref{thm:opt-sol}.
\begin{theorem}\label{thm:opt-schedule}
If $\bar\omega>\frac{N_0W}{\alpha}$, then $\theta_{\rm ef}^*$
is given by
$$\theta_{\rm ef}^*:~\{\omega_k
=\frac{N_0W}{\alpha n_\tau}
\varepsilon_k'\Sigma_\tau^{-1}\varepsilon_k
+\bar \omega-\frac{N_0W}{\alpha}\},$$
where
$\Sigma_{\tau+1}=\frac{n_\tau}{n_\tau+2}A
\Sigma_{\tau}A'+h(\overline P)-\overline P$ with $\Sigma_0=0$.
Otherwise, if $\bar \omega\leq \frac{N_0W}{\alpha},$
$\theta_{\rm ef}^*$ is given by
$$\theta_{\rm ef}^*:~\{\omega_k
=\frac{\bar \omega}{n_\tau}
\varepsilon_k'\Sigma_\tau^{-1}\varepsilon_k
\},$$
where $\Sigma_{\tau+1}=
\frac{n_\tau N_0W}{n_\tau N_0W+2\alpha
\bar \omega}A
\Sigma_{\tau}A'+h(\overline P)-\overline P$.
\end{theorem}
\begin{remark}\label{remark:better-performance}
{A non-data-driven baseline power controller
with a constant power level $\bar \omega$ is
feasible to
Problem~\ref{problem:optimization_online}. Since
$\theta_{\rm ef}^*$ is the optimal solution, it has not
worse state estimation performance
compared with the alternative non-data-driven
power controller. Numerical examples in
Section~\ref{sec:simulation} demonstrate performance improvements using
$\theta_{\rm ef}^*$ compared with the non-data-driven power controller.}
\end{remark}
The following proposition shows that the rank of $\Sigma_\tau$ 
can be calculated offline. The proof is given in the Appendix.
\begin{proposition}\label{proposition: rank}
Consider the $\theta_{\rm ef}^*$ given in Theorem~\ref{thm:opt-schedule},
for any $\tau\in\mathbb N_+$, $n_\tau$ can be calculated
as:
$n_\tau=\mathrm{rank}(h^\tau(\overline{P})-\overline{P}). $
In particular, when $\tau\geq n$, the dimension of $x$, $n_\tau$
becomes a constant which is given by:
$n_\tau=\mathrm{rank}(h^{n}(\overline{P})-\overline{P}),~\forall~
\tau\geq n.$
\end{proposition}

\subsection{Extension 
}\label{subsec:vary-power-in-time}
 In many
cases, the base-line power controller
changes over time with
respect to different
settings. For example, in~\cite{queahl10}, block fading channels were taken
into account. To deal with a time-varying channel power gain $h_k$\footnote{
The term ``channel power gain" means the square of the magnitude of the complex channel.}, a predictive power control algorithm
was established, which
determines the transmission power level, bit rates and
codebooks used by the sensors.
{
The algorithm in~\cite{queahl10} requires
that the receiver (i.e, the remote estimator) runs a channel gain predictor, see e.g.,~\cite{linahl02}.}
{A key observation is that the data-driven controller proposed in the present work can be readily adapted to situations where the baseline controller provides time-varying power levels $\bar{w}_k$.\footnote{{ Following assumptions commonly made in the literature, see, e.g., \cite{queahl10,quevedo2012kalman}, in the sequel we shall assume that the channel gain $h_k$ is available via the one-step ahead channel gain predictor.}}}
In fact, by solving Problem~\ref{problem:transform_optimization_online}
for a time-varying power
baseline $\bar \omega_k$, we obtain the optimal
solution $\theta_{\rm ef}^*$ as
follows:~If $\bar\omega_k>\frac{N_0W}{\alpha}$, then $\theta_{\rm ef}^*$
is given by
$$\theta_{\rm ef}^*:~\{\omega_k
=\frac{N_0W}{\alpha n_\tau}
\varepsilon_k'\Sigma_k^{-1}\varepsilon_k
+\bar \omega_k-\frac{N_0W}{\alpha}\}$$
and
$\Psi_{k}=\frac{n_\tau}{n_\tau+2}
\Sigma_{k-1}$.
Otherwise, if $\bar \omega_k\leq \frac{N_0W}{\alpha},$
$\theta_{\rm ef}^*$ is given by
$$\theta_{\rm ef}^*:~\{\omega_k
=\frac{\bar \omega}{n_\tau}
\varepsilon_k'\Sigma_k^{-1}\varepsilon_k
\}$$
and $\Psi_{k}=
\frac{n_\tau N_0W}{n_\tau N_0W+2\alpha
\bar \omega}
\Sigma_{k}.$ In both cases, $\Sigma_k=(1-\gamma_{k-1})A\Psi_{k-1}A'+
h(\overline{P})-\overline{P}$.
Note that $\Sigma_k$, $\Psi_k$ and $\Phi_k$ are calculated similar to $\Sigma_\tau$,
$\Psi_\tau$ and $\Phi_\tau$ given in Theorem~\ref{thm:opt-schedule}.
To reduce the sensor's computational load, the sensor only needs to
calculate the quadratic form $\varepsilon_k'\Sigma_\tau^{-1}\varepsilon_k$,
while the rest of the paraments are updated and then sent to the sensor by the estimator.
Note that calculating $\varepsilon_k'\Sigma_\tau^{-1}\varepsilon_k$
has a complexity of $O(n^2)$.}

\section{Simulation and Examples}
\label{sec:simulation}
{
Consider a system with parameters as follows:
$
A=\left[
      \begin{array}{cc}
        0.99 & 0.3 \\
        0.1 & 0.7\\
      \end{array}
    \right],
C=\left[
      \begin{array}{cc}
        2.3 & 1 \\
        1 & 1.8\\
      \end{array}
    \right],
R=Q=I_{2\times 2}.
$
}
{We first assume that $\theta$
has a constant power baseline
$\bar\omega=5$ and $\frac{N_0W}{\alpha}=3<\bar\omega.$
In Section~\ref{subsec:Comparison under Time-varying channel power gains},
a time-varying power baseline is considered.}

\begin{figure}[thp]
  \centering
  \includegraphics[width=7cm]{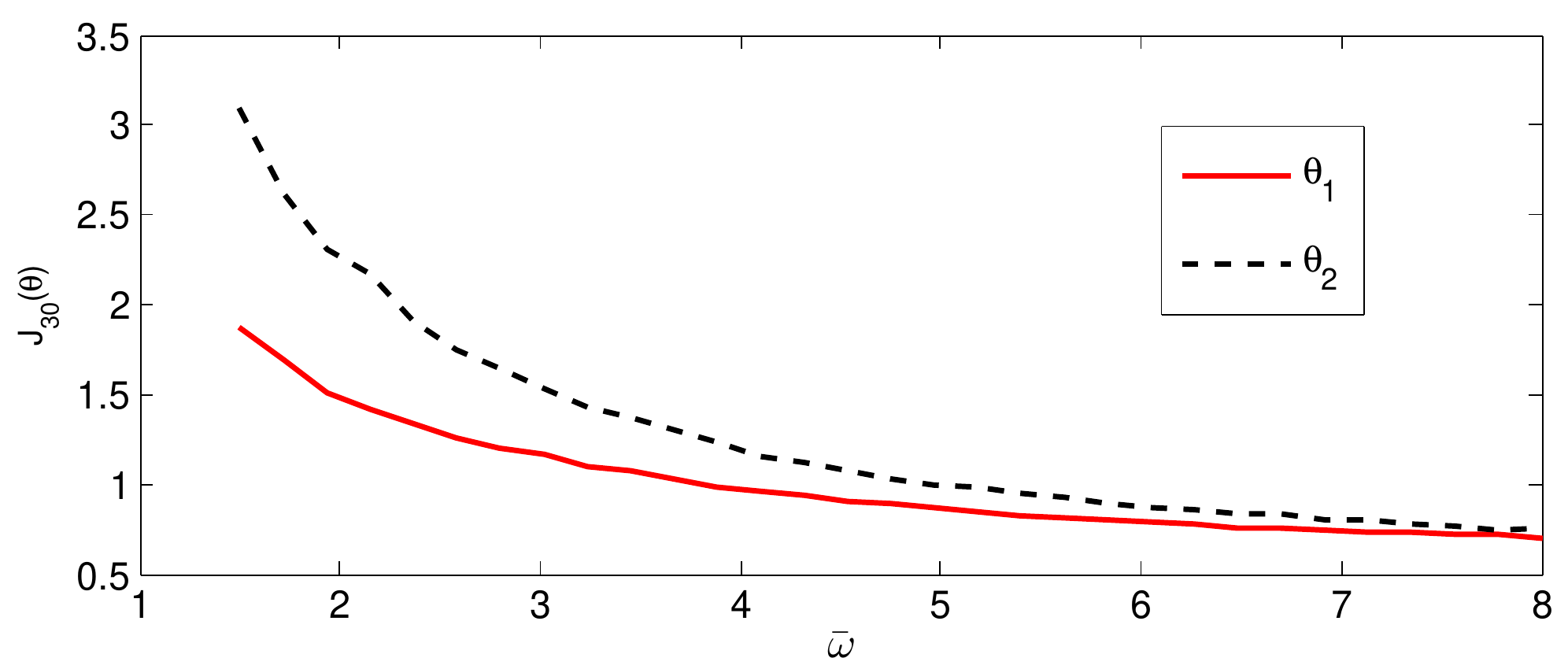}
  \caption{Empirical estimation covariance provided by  controllers $\theta^*_{\text{ef}}(\theta_1)$ and $\theta_2$ as a function of energy constraint $\bar\omega$.} \label{fig:simulation_2}
  \vspace{-1mm}
\end{figure}

\subsection{Comparison with Different Energy Constraints}
We compare our proposed schedule
$\theta_{\rm ef}^*$ (denoted as $\theta_1$)
with a constant baseline power controller
within the entire time horizon (denoted as $\theta_2: \{\omega_k=\bar\omega\}$).
Define
$
  J_k(\theta)=\frac{1}{k}\sum_{i=1}^{k}\mathrm{Tr} \left(\mathbb{E}[P_i]\right)
$
as the empirical approximation (via 100000 Monte Carlo simulations) of the average expected state error covariance (denoted as $J(\theta)$). We choose $J_{30}(\theta)$ as an approximation of $J(\theta)$.

Fig. \ref{fig:simulation_2} shows that $\theta_{\rm ef}^*$ leads to a better system performance when compared to $\theta_2$ under the same energy constraint.

\subsection{Comparison under Fading Channels }\label{subsec:Comparison under Time-varying channel power gains}
In practice, wireless communication channels
typically comprise fading often assumed to be Rayleigh \cite{rappaport1996wireless}, i.e., the channel power gain $h_k$ is exponentially distributed with $\mathrm{pdf}(h_k)=\frac{1}{\overline h}\exp{(-\frac{h_k}{\overline h})},$
where $h_k\geqslant0$ and $\overline h$ is the mean of $h_k$.
{Truncated channel inversion transmit
power controllers have been studied in several works
\cite{quevedo2012kalman,leong2012power,goldsmith1997capacity}, where the transmission power is the inversion of $h_k$,
with a truncated boundary. In this subsection, we use the baseline power
determined by truncated channel gain inversion }
Denote the truncated channel inversion transmission power controller as $\theta_3$:
\begin{equation}\label{eqn:truncated-channel-inversoin-trans-power-controller}
 \omega_k= \left\{\begin{array}{ll}
                                  \frac{v}{h_k}          , &  h_k>h^\star,\\
                                       \frac{v}{h^\star}      ,  & \mathrm{otherwise}.
\end{array}\right.
\end{equation}
where $v$ and $h^\star$ are design parameters.
Consider the case of $\overline h=1$ and set $h^\star=5$. Based on the results in~\cite{leong2012power}, we can choose $v$ to meet the energy constraint.
{
Fig.~\ref{fig:simulation_3} suggests that $\theta_{\rm ef}^*$ leads
to better system performance when compared with $\theta_3$. Fig.~\ref{fig:simulation_4}
shows the comparison given a specific realization of channel power gains.
\begin{figure}[thp]
  \centering
  \includegraphics[width=7cm]{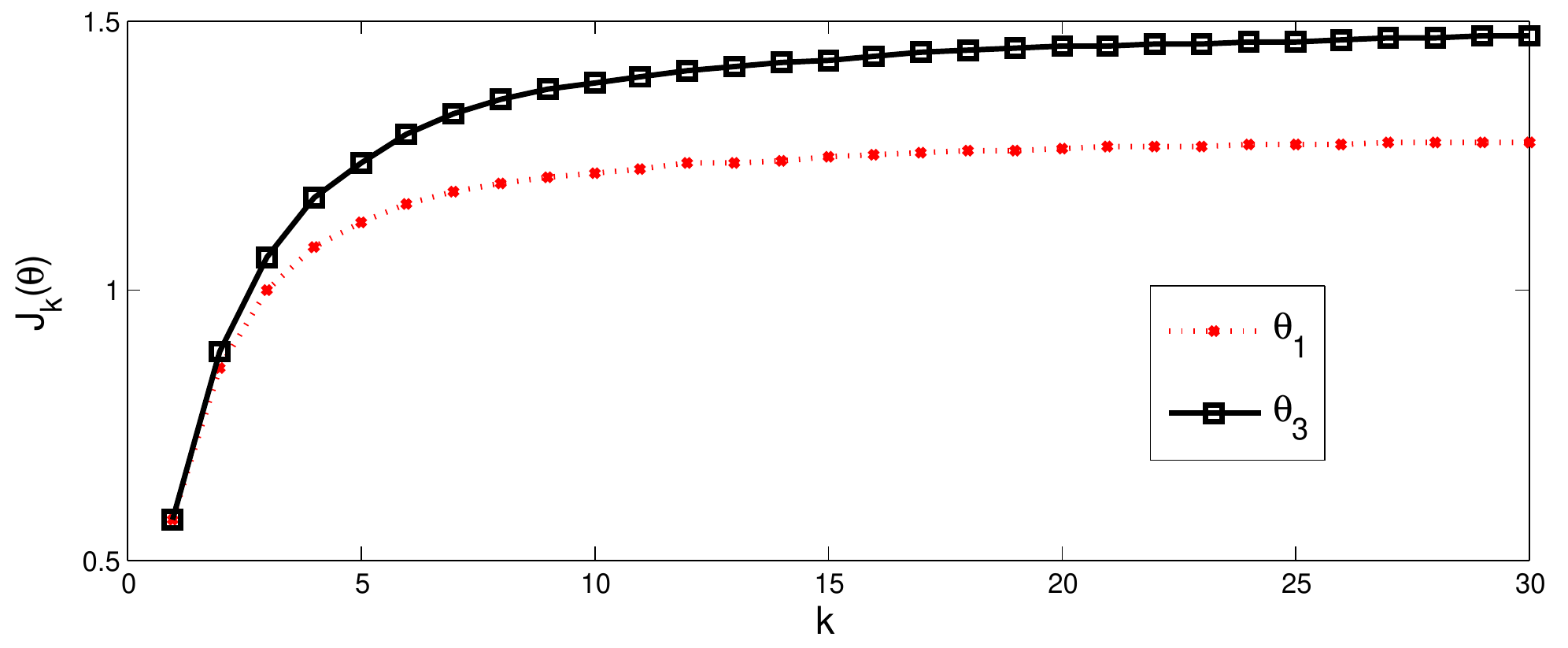}
  \caption{Comparison of $\theta^*_{\text{ef}}(\theta_1)$ and $\theta_3$ under Rayleigh fading.} \label{fig:simulation_3}
  \vspace{-1mm}
\end{figure}
\begin{figure}[thp]
  \centering
  \includegraphics[width=7cm]{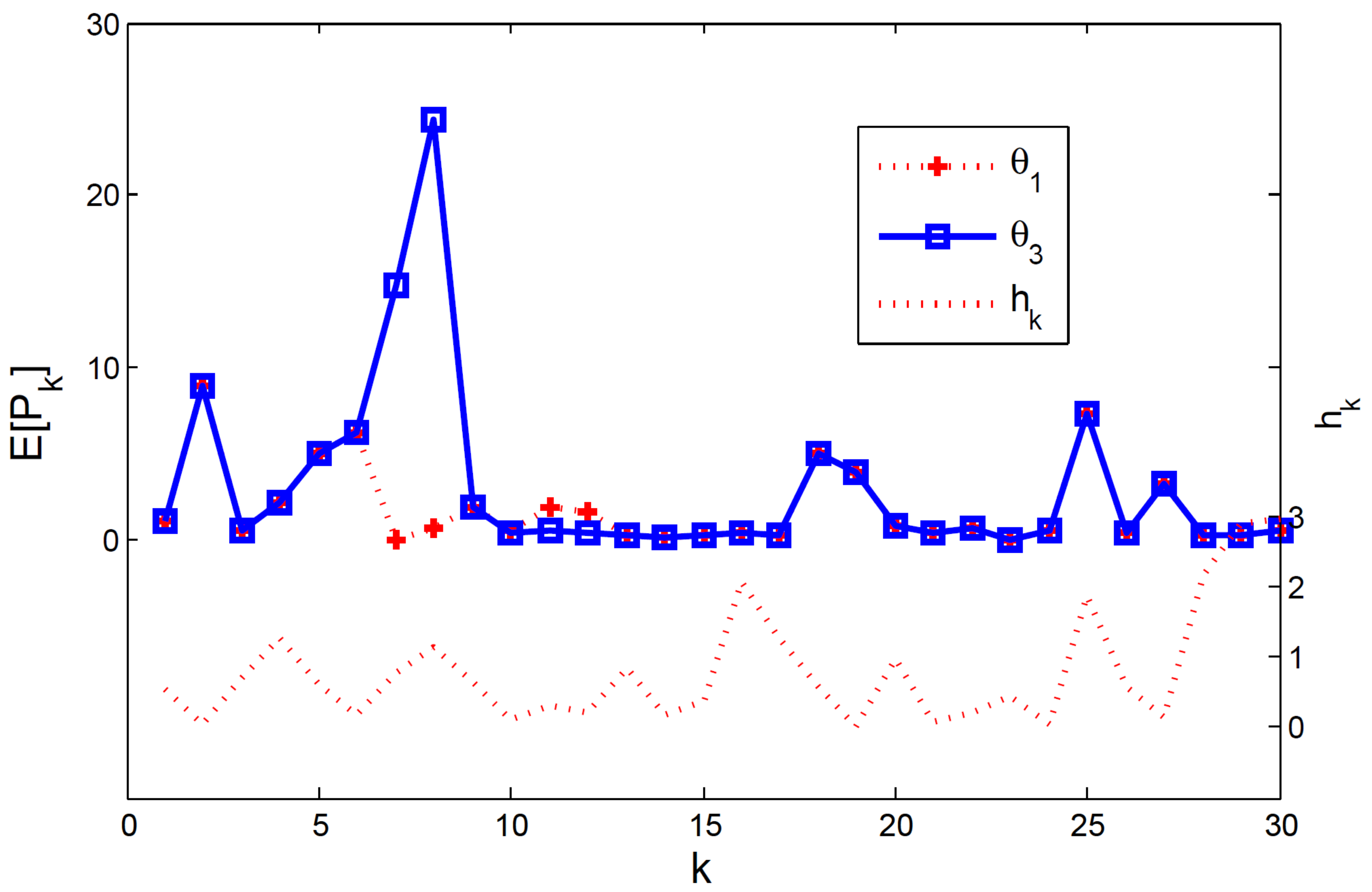}
  \caption{Comparison of $\theta^*_{\text{ef}}(\theta_1)$ and $\theta_3$ given a specific realization of channel power gains.} \label{fig:simulation_4}
  \vspace{-1mm}
\end{figure}

\section{Conclusion}\label{sec:conculsion}
We proposed a data-driven transmission power controller for remote state
estimation, which adjusts the sensor's transmission power according to
its real-time measurements. Then we proved that the proposed power
 controller preserves Gaussianity of the incremental innovation and provided a closed-form
expression of the expected state estimation error covariance.
a tuning method for parameter design was presented to guarantee
that the data-driven power controller
not worse performance than the
alternative non-data-driven ones.
Comparisons were conducted to illustrate estimation performance improvement.

\section*{Appendix}


\textit{Proof of Lemma~\ref{lemma:rank-Phi}:~}
To verify the clain, it
suffices to show that $\mathrm{rank}(\Phi_\tau
)\geq n_\tau$.
Suppose that $\mathrm{rank}(\Phi_\tau
)=r<n_\tau$. Since $\Phi_\tau\succeq 0$,
there must exist exactly $n-r$ mutually orthogonal vectors
$\mathbf{e}_1,\ldots,\mathbf{e}_{n-r}$
such that
${\mathbf{e}_i}'\Phi_\tau\mathbf{e}_i=0,\hbox{~for~}i=1,\ldots,
n-r.$
Denote the unit vector with
only the $(n_\tau\hspace{-0.8mm}+\hspace{-0.5mm}j)$th
entry being $1$ by
$\mathbf{i}_j$, that is,
$\mathbf{i}_j=[\,\underbrace{0,\ldots,0,1}
_{n_\tau+j},0,\ldots,0\,]'.$
Since
${\mathbf{i}_j}'\Phi_\tau\mathbf{i}_j=0,~
j
=1,\ldots,
n-n_\tau,$ without loss of generality, let
$\mathbf{e}_j=\mathbf{i}_{j}$.
As we assume that $\mathbf{e}_{n-r}$ is orthogonal to
$\mathbf{e}_j,~j
=1,\ldots,
n-n_\tau,$
it is true that $D_\tau^{1/2}\mathbf{e}_{n-r}\neq 0$.
Since $U_\tau$ is nonsingular and $\mathrm{Ker}(U_\tau)
=\{0\}$,
we have $\mathbf{e}\triangleq\Sigma_\tau^{1/2}\mathbf{e}_{n-r}\neq 0$.
We then observe that
$\mathbf{e}'\Psi_\tau^{-1}\mathbf{e}=
{\mathbf{e}_{n-r}}'\Phi_\tau\mathbf{e}
_{n-r}=0,$
and
$
\mathbf{e}'\Sigma_\tau^{-1}\mathbf{e}=
{\mathbf{e}_{n-r}}'\left[
\begin{array}{cc}
I_{n_\tau} & 0\\
0& 0\end{array}\right]\mathbf{e}_{n-r}>0,
$
which contradicts with $\Psi_\tau^{-1}\succeq
\Sigma_\tau^{-1}$.
\hfill$\blacksquare$

\textit{Proof of Lemma~\ref{lemma:det-transform}:~}
By definition, it is easy to see that
$\mathrm{det}(\Sigma_\tau)\mathrm{det}
(\Psi_\tau^{-1})=\prod_{i=1}^{n_\tau}
\lambda_i(\Sigma_\tau){\lambda_i(\Psi_\tau)}^{-1}.$
Therefore we only need to prove
$\mathrm{det}(\Phi_\tau)=\prod_{i=1}^{n_\tau}
\lambda_i(\Sigma_\tau){\lambda_i
(\Psi_\tau)}^{-1}.$
Observe that
$\Sigma_\tau$ and $\Psi_\tau$ can be factorized
as
$\Sigma_\tau=U_\tau \left[
\begin{array}{cc}
\Delta_\tau & 0\\
0 &0\end{array}\right]{U_\tau}'$ and
$\Psi_\tau=V_\tau \left[
\begin{array}{cc}
\Theta_\tau & 0\\
0&0\end{array}\right]{V_\tau}'$,
where
$\Delta_\tau$ and $\Theta_\tau$
are diagonal matrices generated
respectively
by the nonzero eigenvalues of $\Sigma_\tau$ and
$\Psi_\tau$. For $i=1,\ldots, n_\tau$,
$u_i$ and $v_i$ are the eigenvectors
associated with
$\lambda_i(\Sigma_\tau)$ and
$\lambda_i(\Psi_\tau)$. 
In addition,
$U_\tau=[\,{u}_1,\ldots,
{u}_{n_\tau},{0},\ldots,
{0}\,]$
and $V_\tau=[\,{v}_1,\ldots,
{v}_{n_\tau},{0},\ldots,
{0}\,]$. Then $\Phi_\tau$ can be
written as
$
\Phi_\tau=\left[
\begin{array}{cc}
M_\tau & 0\\
0&0
\end{array}\right],
$
where $M_\tau\hspace{-1mm}=\hspace{-1mm}{\Delta_\tau}^{1/2}
{\tilde{U}_\tau}'
\tilde{V}_\tau
{\Theta_\tau}^{-1}
{\tilde{V}_\tau}'
\tilde{U}_\tau
{\Delta_\tau}^{1/2}\hspace{-1mm}\in \hspace{-0.5mm}\mathbb{S}_+^{n_\tau}$,
$\tilde{U}_\tau=[\,u_1,\ldots,u_{n_\tau}]$ and
$\tilde{V}_\tau=[\,v_1,\ldots,v_{n_\tau}]$.
According to Lemma 4.3,
$M_\tau$ is nonsingular, so
$\mathrm{det}(\Phi_\tau)=\mathrm{det}
(M_\tau)$.
Since $\mathrm{Im}(\Sigma_\tau)=
\mathrm{Im}(\Psi_\tau)$ from~\eqref{eqn:same-image},
there exists a
unitary matrix $V$ such that $\tilde{V}_\tau=
\tilde{U}_\tau V$. Thus,
$
\mathrm{det}(M_\tau)=\mathrm{det}
\left({\Delta_\tau}^{1/2}V{\Theta_\tau}^{-1}
V'{\Delta_\tau}^{1/2}\right)=\mathrm{det}
\left(
{\Delta_\tau}{\Theta_\tau}^{-1}\right),
$
which completes the proof. \hfill $\blacksquare$

\textit{Proof of Lemma~\ref{thm:opt-sol}:~}
According to Lemma~\ref{lemma:equal-eig}, we
set $\lambda_1(\tau)=\cdots=\lambda_{n_\tau}(\tau)=
\lambda_\tau$. Logarithm does not change the
monotonicity of~\eqref{eqn:objective-func-opt}.
Problem~\ref{problem:transform_optimization_online}
is consequently transformed to\vspace{-2mm}
\begin{eqnarray}
&\min\limits_{\lambda_\tau,\omega}&-\frac{\alpha}{N_0W}\omega
-(\frac{n_\tau}{2}+1)\ln{\lambda_\tau},\label{eqn:derivative}\\
&\hbox{s.t.}&\frac{n_\tau N_0W}{2\alpha}(\lambda_\tau-1)+\omega=\bar
\omega,~~ \omega\geq 0.\notag
\end{eqnarray}
Substituting $\omega=-\frac{\alpha}{N_0W}\bar \omega
+\frac{n_\tau}{2}(\lambda_\tau-1)-(\frac{n_\tau}{2}
+1)\ln{\lambda_\tau}$ into ~\eqref{eqn:derivative}
and taking derivative, it yields that the minimum
of~\eqref{eqn:derivative} is attained at
$\lambda_\tau=1+\frac{2}{n_\tau}$. Meanwhile
$\omega$ needs to be nonnegative, so
the optimal solution to Problem~\ref{problem:optimization_online} is
\eqref{eqn:opt_sol1} if $\bar\omega>\frac{N_0W}{\alpha}$ or \eqref{eqn:opt_sol2} otherwise.
\hfill $\blacksquare$

\textit{Proof of Proposition~\ref{proposition: rank}:~}
Consider a matrix $\Sigma=\sum_{i=1}^{\tau}
\rho_i\left(
h^{i}(\overline P)-h^{i-1}(\overline{P})\right)$ with
$\rho_i\in(0,1]$.
We have
\begin{eqnarray}\label{eq:Image-eqn}
\mathrm{Im}(\Sigma)&=&\mathrm{Im}([\,\rho_1\Sigma_1^{1/2}~\rho_2A\Sigma_1^{1/2}~
\cdots~
\rho_\tau A^{\tau-1}\Sigma_1^{1/2}\,][\,\cdot\,]')\notag\\
&=&\mathrm{Im}([\,\rho_1\Sigma_1^{1/2}~\rho_2A\Sigma_1^{1/2}~\cdots~
\rho_\tau A^{\tau-1}\Sigma_1^{1/2}\,])\notag\\
&=&\mathrm{Im}([\,\Sigma_1^{1/2}~A\Sigma_1^{1/2}~\cdots~
A^{\tau-1}\Sigma_1^{1/2}\,])\notag\\
&=&\mathrm{Im}([\,\Sigma_1^{1/2}~A\Sigma_1^{1/2}~\cdots~
A^{\tau-1}\Sigma_1^{1/2}\,][\,\cdot\,]')\notag\\
&=&\mathrm{Im}(h^\tau(\overline P)-\overline P),
\end{eqnarray}
which leads to the first assertion.
By the Cayley-Hamilton theorem,
we have
$A^k=-a_1(k)A^{n-1}-a_2(k)A^{n-2}-\cdots-
a_n(k)I,~~\forall~k\geq n,$
where $a_1(k),\ldots,a_n(k)$ are coefficients of the
characteristic polynomial of $A$.
When $\tau\geq n+1$, we have
\begin{eqnarray*}
&&\mathrm{Im}([\,\Sigma_1^{1/2}~A\Sigma_1^{1/2}~\cdots~
A^{\tau-1}\Sigma_1^{1/2}\,][\,\cdot\,]')\\
&=&
\mathrm{Im}([\,\Sigma_1^{1/2}~A\Sigma_1^{1/2}~\cdots~
-a_1(\tau\hspace{-1mm}-\hspace{-1mm}1)A^{n-
1}\Sigma_1^{1/2}\\
&&\;\;\;\;\;\;\;-a_2(\tau\hspace{-1mm}-\hspace{-1mm}1)
A^{n-2}\Sigma_1^{1/2}-\cdots-
a_n(\tau\hspace{-1mm}-\hspace{-1mm}1)
\Sigma_1^{1/2}\,]),
\end{eqnarray*}
The last assertion follows from the reasoning used in~\eqref{eq:Image-eqn}.
\hfill $\blacksquare$

\bibliographystyle{IEEETran}
\bibliography{reference1,dquevedo,sj_reference}

\begin{thebibliography}{10}
\providecommand{\url}[1]{#1}
\csname url@samestyle\endcsname
\providecommand{\newblock}{\relax}
\providecommand{\bibinfo}[2]{#2}
\providecommand{\BIBentrySTDinterwordspacing}{\spaceskip=0pt\relax}
\providecommand{\BIBentryALTinterwordstretchfactor}{4}
\providecommand{\BIBentryALTinterwordspacing}{\spaceskip=\fontdimen2\font plus
\BIBentryALTinterwordstretchfactor\fontdimen3\font minus
  \fontdimen4\font\relax}
\providecommand{\BIBforeignlanguage}[2]{{%
\expandafter\ifx\csname l@#1\endcsname\relax
\typeout{** WARNING: IEEEtran.bst: No hyphenation pattern has been}%
\typeout{** loaded for the language `#1'. Using the pattern for}%
\typeout{** the default language instead.}%
\else
\language=\csname l@#1\endcsname
\fi
#2}}
\providecommand{\BIBdecl}{\relax}
\BIBdecl

\bibitem{shi2012optimal}
L.~Shi and L.~Xie, ``{Optimal sensor power scheduling for state estimation of
  Gauss--Markov systems over a packet-dropping network},'' \emph{IEEE
  Transactions on Signal Processing}, vol.~60, no.~5, pp. 2701--2705, 2012.

\bibitem{xu2004optimal}
Y.~Xu and J.~P. Hespanha, ``Optimal communication logics in networked control
  systems,'' in \emph{Proceedings of the 43rd IEEE Conference on Decision and
  Control}, vol.~4.\hskip 1em plus 0.5em minus 0.4em\relax IEEE, 2004, pp.
  3527--3532.

\bibitem{Imer05CDC}
O.~C. Imer and T.~Basar, ``Optimal estimation with limited measurements,'' in
  \emph{Proceedings of the 44th IEEE Conference on Decision and Control,
  European Control}, December 2005, pp. 1029--1034.

\bibitem{queahl10}
D.~E. Quevedo, A.~Ahl\'{e}n, and J.~{\O stergaard}, ``Energy efficient state
  estimation with wireless sensors through the use of predictive power control
  and coding,'' \emph{{IEEE} Transactions Signal Processing}, vol.~58, no.~9,
  pp. 4811--4823, 2010.

\bibitem{leong2012power}
A.~S. Leong and S.~Dey, ``{Power allocation for error covariance minimization
  in Kalman filtering over packet dropping links},'' in \emph{Decision and
  Control (CDC), 2012 IEEE 51st Annual Conference on}.\hskip 1em plus 0.5em
  minus 0.4em\relax IEEE, 2012, pp. 3335--3340.

\bibitem{nouleo14}
M.~Nourian, A.~Leong, S.~Dey, and D.~E. Quevedo, ``An optimal transmission
  strategy for {K}alman filtering over packet dropping links with imperfect
  acknowledgements,'' \emph{IEEE Trans. Contr. Network Syst.}, vol.~1, no.~3,
  pp. 259--271, Sept. 2014.

\bibitem{quevedo2012kalman}
D.~E. Quevedo, A.~Ahl{\'e}n, A.~S. Leong, and S.~Dey, ``{On Kalman filtering
  over fading wireless channels with controlled transmission powers},''
  \emph{Automatica}, vol.~48, no.~7, pp. 1306--1316, 2012.

\bibitem{GatsisACC13}
K.~Gatsis, A.~Ribeiro, and G.~J. Pappas, ``Optimal power management in wireless
  control systems,'' in \emph{American Control Conference (ACC), 2013}, 2013,
  pp. 1562--1569.

\bibitem{box2011bayesian}
G.~E. Box and G.~C. Tiao, \emph{Bayesian inference in statistical
  analysis}.\hskip 1em plus 0.5em minus 0.4em\relax Wiley-Interscience, 2011.

\bibitem{Liyuzhe13CDC}
Y.~Li, D.~E. Quevedo, V.~Lau, and L.~Shi, ``Online sensor transmission power
  schedule for remote state estimation,'' in \emph{Proceedings of 52nd IEEE
  Conference on Decision and Control}, Florence, Italy, 2013.

\bibitem{hovareshti2007sensor}
P.~Hovareshti, V.~Gupta, and J.~S. Baras, ``Sensor scheduling using smart
  sensors,'' in \emph{Proceedings of the 46th IEEE Conference on Decision and
  Control}, 2007, pp. 494--499.

\bibitem{andmoo79}
B.~D.~O. Anderson and J.~Moore, \emph{Optimal Filtering}.\hskip 1em plus 0.5em
  minus 0.4em\relax Englewood Cliffs, NJ: Prentice Hall, 1979.

\bibitem{sinopoli2004kalman}
B.~Sinopoli, L.~Schenato, M.~Franceschetti, K.~Poolla, M.~I. Jordan, and S.~S.
  Sastry, ``Kalman filtering with intermittent observations,'' \emph{IEEE
  Transactions on Automatic Control}, vol.~49, no.~9, pp. 1453--1464, 2004.

\bibitem{fu2009Automatica}
M.~Fu and C.~E. de~Souza, ``State estimation for linear discrete-time systems
  using quantized measurements,'' \emph{Automatica}, vol.~45, no.~12, pp. 2937
  -- 2945, 2009.

\bibitem{wu2013event}
J.~Wu, Q.~shan Jia, K.~H. Johansson, and L.~Shi, ``Event-based sensor data
  scheduling: Trade-off between communication rate and estimation quality,''
  \emph{IEEE Transactions on Automatic Control}, vol.~58, no.~4, pp.
  1041--1046, 2013.

\bibitem{kotecha2003gaussian}
J.~H. Kotecha and P.~M. Djuric, ``Gaussian particle filtering,'' \emph{IEEE
  Transactions on Signal Processing}, vol.~51, no.~10, pp. 2592--2601, 2003.

\bibitem{soi-kf-tsp06}
A.~Ribeiro, G.~B. Giannakis, and S.~I. Roumeliotis,
  ``$\mathrm{SOI}$-$\mathrm{KF}$: Distributed $\mathrm{K}$alman filtering with
  low-cost communications using the sign of innovations,'' \emph{IEEE
  Transactions on Signal Processing}, vol.~54, no.~12, pp. 4782--4795, 2006.

\bibitem{linahl02}
L.~Lindbom, A.~Ahl\'{e}n, M.~Sternad, and M.~Falkenstr\"om, ``Tracking of
  time-varying mobile radio channels--part {II}: A case study,'' \emph{{IEEE}
  Transactions Commun.}, vol.~50, no.~1, pp. 156--167, Jan. 2002.

\bibitem{rappaport1996wireless}
T.~S. Rappaport \emph{et~al.}, \emph{Wireless communications: principles and
  practice}.\hskip 1em plus 0.5em minus 0.4em\relax Prentice Hall PTR New
  Jersey, 1996, vol.~2.

\bibitem{goldsmith1997capacity}
A.~J. Goldsmith and P.~P. Varaiya, ``Capacity of fading channels with channel
  side information,'' \emph{IEEE Transactions on Information Theory}, vol.~43,
  no.~6, pp. 1986--1992, 1997.

\end{thebibliography}

\end{document}